\documentclass[prd,aps,floatfix,nofootinbib,11 pt]{revtex4}
\usepackage{amssymb}
\usepackage{amsmath,graphicx,color,epsfig}

\input{tcilatex}

\begin{document}

\title{Entanglement dynamics of non-inertial observers in a correlated
environment}
\author{M. Ramzan\thanks{%
mramzan@phys.qau.edu.pk}}
\address{Department of Physics Quaid-i-Azam University \\
Islamabad 45320, Pakistan}

\date{\today }

\begin{abstract}
Effect of decoherence and correlated noise on the entanglement of $X$-type
state of the Dirac fields in the non-inertial frame is investigated. A two
qubit $X$-state is considered to be shared between the partners where Alice
is\ in inertial frame and Rob in an accelerated frame. The concurrence is
used to quantify the entanglement of the $X$-state system influenced by time
correlated amplitude damping, depolarizing and bit flip channels. It is seen
that amplitude damping and bit flip channels heavily influence the
entanglement of the system as compared to the depolarizing channel. It is
found possible to avoid entanglement sudden death (ESD) for all the channels
under consideration for $\mu >0.75$ for any type of initial state. No ESD
behaviour is seen for depolarizing channel in the presence of correlated
noise for entire range of decoherence parameter $p$ and Rob's acceleration $r
$. It is also seen that the effect of environment is much stronger than that
of acceleration of the accelerated partner. Furthermore, it is investigated
that correlated noise compensates the loss of entanglement caused by the
Unruh effect.\newline
\end{abstract}

\maketitle

\section{Introduction}

Quantum entanglement is the major resource in quantum information science
and can be used as a potential source for quantum teleportation of unknown
states [1], quantum key distribution [2], quantum cryptography [3] and
quantum computation [4, 5]. Entanglement sudden death for bipartite and
multipartite systems has been the main focus of researchers during recent
years [6-11]. Another important feature, entanglement sudden birth (ESB) has
also been investigated where the initially unentangled qubits can be
entangled after a finite evolution of time [12-13]. Recently, entanglement
behavior in non-inertial frames was investigated by Alsing et al. [14]. They
studied the fidelity of teleportation between relative accelerated partners.
Quantum information in a relativistic setup has become an interesting topic
of research during recent years [15-27].

Since, quantum systems are influenced by their environment that may results
in the non-unitary dynamics of the system. Therefore, the environmental
effect on a quantum system gives rise to the phenomenon of decoherence that
causes an irreversible transfer of information from the system to the
environment [28-29]. Study of decoherence in non-inertial frames have been
investigated for bipartite and multipartite systems by number of authors
[30-33], where it is shown that entanglement is degraded by the acceleration
of the inertial observers. Recently, a qubit-qutrit system in non-inertial
frames has been analyzed under decoherence [34], where it is shown that ESB
does occur in case of depolarizing channel. On the other hand, quantum
channels with memory [35-39] provide a natural theoretical framework for the
study of any noisy quantum communication. The main focus of this work is to
study the entanglement dynamics in the presence of correlated noise as it
has not been studied yet in non-inertial frames.

In this paper, decoherence and correlated noise effects are investigated for
$X$-type states in non-inertial frames by considering using amplitude
damping, depolarizing and bit flip channels. The two observers Alice and Rob
share an $X$-type state in non-inertial frames. Alice is considered to be
stationary whereas Rob moves with a uniform acceleration. Two important
features of entanglement, ESD and ESB are investigated. No ESD occurs in
case of depolarizing channel in the presence of correlated noise.

\section{Open system dynamics of non-inertial observers under correlated
noise}

The evolution of a system and its environment can be described by
\begin{equation}
U_{SE}(\rho _{S}\otimes |0\rangle _{E}\left\langle 0\right\vert
)U_{SE}^{\dag }
\end{equation}%
where $U_{SE}$ represents the evolution operator for the combined system and
$|0\rangle _{E}$ corresponds to the initial state of the environment. By
taking trace over the environmental degrees of freedom, the evolution of the
system can be obtained as%
\begin{eqnarray}
L(\rho _{S}) &=&\text{Tr}_{E}\{U_{SE}(\rho _{S}\otimes |0\rangle
_{E}\left\langle 0\right\vert )U_{SE}^{\dag }\}  \notag \\
&=&\sum\nolimits_{\mu }\ _{E}\left\langle \mu \right\vert U_{SE}|0\rangle
_{E}\rho _{SE}\left\langle 0\right\vert )U_{SE}^{\dag }|\mu \rangle _{E}
\end{eqnarray}%
where $|\mu \rangle _{E}$ represents the orthogonal basis of the environment
and $L$ is the operator describing the evolution of the system. The above
equation can also be written as%
\begin{equation}
L(\rho _{S})=\sum\nolimits_{\mu }M_{\mu }\rho _{S}M_{\mu }^{\dag }
\end{equation}%
where $M_{\mu }=\ _{E}\left\langle \mu \right\vert U_{SE}|0\rangle _{E}$ are
the Kraus operators as given in Ref. [40]. The Kraus operators satisfy the
completeness relation%
\begin{equation}
\sum\nolimits_{\mu }M_{\mu }^{\dag }M_{\mu }=1
\end{equation}%
The decoherence process can also be represented by a map in terms of the
complete system-environment state. The dynamics of a $d$-dimensional quantum
system can be represented by the following map [41]%
\begin{equation}
U_{SE}|\xi _{l}\rangle _{S}\otimes |0\rangle _{E}=\sum\nolimits_{k}M_{k}|\xi
_{l}\rangle _{S}\otimes |k\rangle _{E}
\end{equation}%
where $\{|\xi _{l}\rangle _{S}\}$ $(l=1,.....,d)$ is the complete basis for
the system and%
\begin{eqnarray}
|\xi _{1}\rangle _{S}\otimes |0\rangle _{E} &\rightarrow &M_{0}|\xi
_{1}\rangle _{S}\otimes |0\rangle _{E}+.....+M_{d^{2}-1}|\xi _{1}\rangle
_{S}\otimes |d^{2}-1\rangle _{E}  \notag \\
|\xi _{2}\rangle _{S}\otimes |0\rangle _{E} &\rightarrow &M_{0}|\xi
_{2}\rangle _{S}\otimes |0\rangle _{E}+.....+M_{d^{2}-1}|\xi _{2}\rangle
_{S}\otimes |d^{2}-1\rangle _{E}  \notag \\
. &&  \notag \\
. &&  \notag \\
. &&  \notag \\
|\xi _{d}\rangle _{S}\otimes |0\rangle _{E} &\rightarrow &M_{0}|\xi
_{d}\rangle _{S}\otimes |0\rangle _{E}+.....+M_{d^{2}-1}|\xi _{d}\rangle
_{S}\otimes |d^{2}-1\rangle _{E}
\end{eqnarray}

Let Alice and Rob (the accelerated observer) share the following $X$-type
initial state%
\begin{equation}
\rho _{AR}=\frac{1}{4}\left( I_{AR}+\sum\limits_{i=0}^{3}c_{i}\sigma
_{i}^{(A)}\otimes \sigma _{i}^{(R)}\right)
\end{equation}%
where $I_{AR}$ is the identity operator in a two-qubit Hilbert space, $%
\sigma _{i}^{(A)}$ and $\sigma _{i}^{(R)}$ are the Pauli operators of the
Alice's and Rob's qubit and $c_{i}$ $(0\leq |c_{i}|\leq 1)$ are real numbers
satisfying the unit trace and positivity conditions of the density operator $%
\rho _{AR}.$ In order to study the entanglement dynamics, different cases
for initial state are considered, for example, the general initial state $%
(|c_{1}|=0.7,|c_{2}|=0.9,|c_{3}|=0.4)$, the Werner initial state $%
(|c_{1}|=|c_{2}|=|c_{3}|=0.8)$, and Bell basis state $%
(|c_{1}|=|c_{2}|=|c_{3}|=1)$.

Let the Dirac fields, as shown in Refs. [42, 43], from an inertial
perspective, can be described by a superposition of Unruh monochromatic
modes $|0_{U}\rangle =\otimes _{\omega }|0_{\omega }\rangle _{U}$ and $%
|1_{U}\rangle =\otimes _{\omega }|1_{\omega }\rangle _{U}$ with
\begin{equation}
|0_{\omega }\rangle _{U}=\cos r|0_{\omega }\rangle _{I}|0_{\omega }\rangle
_{II}+\sin r|1_{\omega }\rangle _{I}|1_{\omega }\rangle _{II}
\end{equation}%
and%
\begin{equation}
|1_{\omega }\rangle _{M}=|1_{\omega }\rangle _{I}|0_{\omega }\rangle _{II}
\end{equation}%
where $\cos r=(e^{-2\pi \omega c/a}+1)^{-1/2}$, $a$ is the acceleration of
the observer, $\omega $ is frequency of the Dirac particle and $c$ is the
speed of light in vacuum. The subscripts $I$ and $II$ of the kets represent
the Rindler modes in region $I$ and $II$, respectively, as shown in the
Rindler spacetime diagram (see Ref. [31], Fig. (1)). By using equations (8)
and (9), equation (7) can be re-written in terms of Minkowski modes for
Alice ($A$) and Rindler modes for Rob $(\tilde{R})$. The single-mode
approximation is used in this study, i.e. a plane wave Minkowski mode is
assumed to be the same as a plane wave Unruh mode (superposition of
Minkowski plane waves with single-mode transformation to Rindler modes).
Therefore, Alice being an inertial observer while her partner Rob who is in
uniform acceleration, are considered to carry their detectors sensitive to
the $\omega $ mode. To study the entanglement in the state from their
perspective one must transform the Unruh modes to Rindler modes. Hence,
Unruh states must be transformed into the Rindler basis. Let Rob detects a
single Unruh mode and Alice detects a monochromatic Minkowski mode of the
Dirac field. Considering that an accelerated observer in Rindler region $I$
has no access to the field modes in the causally disconnected region $II$
and by taking the trace over the inaccessible modes, one obtains the
following density matrix%
\begin{eqnarray}
\rho _{A\tilde{R}} &=&\frac{1}{4}\left(
\begin{array}{cccc}
(1+c_{3})\cos ^{2}r & 0 &  &  \\
0 & (1+c_{3})\sin ^{2}r+(1-c_{3}) &  &  \\
0 & c^{+}\cos r &  &  \\
c^{-}\cos r & 0 &  &
\end{array}%
\right.   \notag \\
&&\left.
\begin{array}{cccc}
&  & 0 & c^{-}\cos r \\
&  & c^{+}\cos r & 0 \\
&  & (1-c_{3})\cos ^{2}r & 0 \\
&  & 0 & (1-c_{3})+(1+c_{3})\sin ^{2}r%
\end{array}%
\right)
\end{eqnarray}%
where $c^{+}=c_{1}+c_{2}$ and $c^{-}=c_{1}-c_{2}.$

Since noise is a major hurdle while transmitting quantum information from
one party to other through classical and quantum channels. This noise causes
a distortion of the information sent through the channel. It is considered
that the system is strongly correlated quantum system, the correlation of
which results from the memory of the channel itself. The action of a Pauli
channel with partial memory on a two qubit state can be written in Kraus
operator formalism as [35]%
\begin{equation}
A_{ij}=\sqrt{p_{i}[(1-\mu )p_{j}+\mu \delta _{ij}]}\sigma _{i}\otimes \sigma
_{j}
\end{equation}%
where $\sigma _{i}$ ($\sigma _{j})$ are usual Pauli matrices, $p_{i}$ ($p_{j}
$) represent the decoherence parameter and indices $i$ and $j$ runs from $0$
to $3.$ The above expression means that with probability $\mu $ the channel
acts on the second qubit with the same error operator as on the first qubit,
and with probability $(1-\mu )$ it acts on the second qubit independently.
Physically the parameter $%
%TCIMACRO{\U{3bc} }%
%BeginExpansion
\mu
%EndExpansion
$ is determined by the relaxation time of the channel when a qubit passes
through it. The action of a two qubit Pauli channel when both the qubits of
Alice and Rob are streamed through it, can be described in operator sum
representation as [46]
\begin{equation}
\rho _{f}=\sum\limits_{k_{1},\text{ }k_{2}=0}^{1}(A_{k_{2}}\otimes
A_{k_{1}})\rho _{in}(A_{k_{1}}^{\dagger }\otimes A_{k_{2}}^{\dagger })
\end{equation}%
where $\rho _{in}$ represents the initial density matrix for quantum state
and $A_{k_{i}}$\ are the Kraus operators as expressed in equation (11). A
detailed list of single qubit Kraus operators for different quantum channels
with uncorrelated noise is given in table 1. Whereas, the Kraus operators
for amplitude damping channel with correlated noise are given by Yeo and
Skeen [36]%
\begin{equation}
A_{00}^{c}=\left[
\begin{array}{llll}
\cos \chi  & 0 & 0 & 0 \\
0 & 1 & 0 & 0 \\
0 & 0 & 1 & 0 \\
0 & 0 & 0 & 1%
\end{array}%
\right] ,\ \ \ A_{11}^{c}=\left[
\begin{array}{llll}
0 & 0 & 0 & 0 \\
0 & 0 & 0 & 0 \\
0 & 0 & 0 & 0 \\
\sin \chi  & 0 & 0 & 0%
\end{array}%
\right]
\end{equation}%
where, $0\leq \chi \leq \pi /2$ and is related to the quantum noise
parameter as
\begin{equation}
\sin \chi =\sqrt{p}
\end{equation}%
The action of such a channel with memory can be written as
\begin{equation}
\pi \rightarrow \rho =\Phi (\pi )=(1-\mu
)\sum\limits_{i,j=0}^{1}A_{ij}^{u}\pi A_{ij}^{u\dagger }+\mu
\sum\limits_{k=0}^{1}A_{kk}^{c}\pi A_{kk}^{c\dagger }
\end{equation}%
where the superscripts $u$ and $c$ represent the uncorrelated and correlated
parts respectively. The Kraus operators are of dimension $2^{2}$ and are
constructed from single qubit Kraus operators by taking their tensor product
over all $n^{2}$ combinations
\begin{equation}
A_{k}=\underset{k_{i}}{\otimes }A_{k_{i}}
\end{equation}%
where $i$ is the number of Kraus operator for a single qubit channel. The
final state of the system after the action of the channel can be obtained as
\begin{equation}
\rho _{f}=\Phi _{p,\mu }(\rho _{A,I})
\end{equation}%
where $\Phi _{p,\mu }$ is the super-operator realizing the quantum channel
parametrized by real numbers $p$ and $\mu $. Since, the entanglement
dynamics of the bipartite subsystems (especially the system-environment
dynamics) is of interest here, therefore, only bipartite reduced matrices
are considered. It is assumed that both Alice and Rob's qubits are
influenced by the time correlated environment. The entanglement of a
two-qubit mixed state $\rho $ in a noisy environment can be quantified by
the concurrence as defined by [45]%
\begin{equation}
C(\rho )=\max \{0,\lambda _{1}-\lambda _{2}-\lambda _{3}-\lambda _{4}\},%
\text{ \ \ \ }\lambda _{i}\geqslant \lambda _{i+1}\geqslant 0
\end{equation}%
where $\lambda _{i}$ are the square roots of the eigenvalues of the matrix $%
\rho _{f}\tilde{\rho}_{f},$ with $\tilde{\rho}_{f}$ being the spin flip
matrix of $\rho _{f}$ and is given by%
\begin{equation}
\tilde{\rho}_{f}=(\sigma _{y}\otimes \sigma _{y})\rho _{f}(\sigma
_{y}\otimes \sigma _{y})
\end{equation}%
where $\sigma _{y}$ is the usual Pauli matrix. Since the density matrix
under consideration has $X$-type structure, therefore a simpler expression
for the concurrence [46] can be used%
\begin{equation}
C(\rho )=2\max \{0,\tilde{C}_{1}(\rho ),\tilde{C}_{2}(\rho )\}
\end{equation}%
where $\tilde{C}_{1}(\rho )=\sqrt{\rho _{14}\rho _{41}}-\sqrt{\rho _{22}\rho
_{33}}$ and $\tilde{C}_{2}(\rho )=\sqrt{\rho _{23}\rho _{32}}-\sqrt{\rho
_{11}\rho _{44}}.$ The reduced-density matrix of the inertial subsystem $A$
and the non-inertial subsystem $\tilde{R}$, can be obtained by taking the
partial trace of $\rho _{A\tilde{R}E_{A}E_{\tilde{R}}}=\rho _{A\tilde{R}%
}\otimes \rho _{A\tilde{R}E_{A}E_{\tilde{R}}}$\ over the degrees of freedom
of the environment i.e.
\begin{equation}
\rho _{A\tilde{R}}=\text{Tr}_{E_{A}E_{\tilde{R}}}\{\rho _{A\tilde{R}E_{A}E_{%
\tilde{R}}}\}
\end{equation}%
which yields the concurrence for the $X$-state structure using equation
(20), under amplitude damping channel as%
\begin{eqnarray}
C^{\text{AD}}(\rho ) &=&\frac{1}{8}\left(
\begin{array}{c}
2\sqrt{c^{+2}(p(\mu -1)+1)^{2}\cos ^{2}(r)} \\
-\sqrt{\left.
\begin{array}{c}
\left. \left(
\begin{array}{c}
2(c_{3}+1)p\mu \cos ^{2}(r)+((p-2)p(\mu -1)-1)\times  \\
(\text{$c_{3}$}+(\text{$c_{3}$}+1)\cos (2r)-3)%
\end{array}%
\right) \times \right.  \\
\left. \left(
\begin{array}{c}
(p+1)(\mu -1)(\text{$c_{3}$}(p-1)+(\text{$c_{3}$}+1)\cos (2r)(p-1) \\
-3p-1)-2(\text{$c_{3}$}+1)(p-1)\mu \cos ^{2}(r)%
\end{array}%
\right) \right.
\end{array}%
\right. }%
\end{array}%
\right)   \notag \\
&&
\end{eqnarray}%
The concurrence of $X$-state system in case of depolarizing channel becomes%
\begin{eqnarray}
C^{\text{Dep}}(\rho ) &=&\frac{1}{16}\left(
\begin{array}{c}
\sqrt{%
\begin{array}{c}
\left. -\cos ^{2}(r)\left(
\begin{array}{c}
(\text{$c_{3}$}+1)p(\text{$c^{-}$}\mu +\text{$c^{+}$}(-4p \\
+(4p-7)\mu +4))\cos ^{2}(r)-2\text{$c^{+}$}(p(\mu -2)+4)%
\end{array}%
\right) \times \right.  \\
\left. \left(
\begin{array}{c}
(\text{$c_{3}$}+1)p(\text{$c^{-}$}\mu +\text{$c^{+}$}(-4p+(4p-7)\mu +4))\cos
^{2}(r) \\
-2\left( \text{$c^{+}$}\left( 4(\mu -1)p^{2}+(6-8\mu )p-4\right) +\text{$%
c^{-}$}p\mu \right)
\end{array}%
\right) \right.
\end{array}%
} \\
-2\sqrt{%
\begin{array}{c}
\left. \cos ^{2}(r)\left(
\begin{array}{c}
\left( -4(\mu -1)p^{2}+8(\mu -1)p+4\right) \text{$c^{-}$}^{2}-\text{$c^{+}$}%
p\mu \text{$c^{-}$} \\
+(\text{$c_{3}$}+1)^{2}(p(\mu -4)+4)\cos ^{2}(r)+4(\text{$c_{3}$}+1)p%
\end{array}%
\right) \right.  \\
\left. \left(
\begin{array}{c}
\frac{1}{4}(\text{$c_{3}$}+1)^{2}(p(\mu -4)+4)\cos ^{4}(r) \\
-\frac{1}{4}\left(
\begin{array}{c}
4\left( (\mu -1)p^{2}-2(\mu -1)p-1\right) \text{$c^{-}$}^{2}+\text{$c^{+}$}%
p\mu \text{$c^{-}$} \\
+4\text{$c_{3}$}(p(\mu -3)+4)+4(p(\mu -3)+4)%
\end{array}%
\right) \times  \\
\cos ^{2}(r)+p(\mu -2)+4%
\end{array}%
\right) \right.
\end{array}%
}%
\end{array}%
\right)   \notag \\
&&
\end{eqnarray}%
and the concurrence of the $X$-state system under the influence of bit flip
channel reads%
\begin{eqnarray}
C^{\text{BF}}(\rho ) &=&\sqrt{\left. \left(
\begin{array}{c}
\text{$c^{+}$}\left( 2(\mu -1)p^{2}-2(\mu -1)p-1\right)  \\
+2\text{$c^{-}$}p(-\mu p+p+\mu -1)%
\end{array}%
\right) ^{2}\cos ^{2}(r)\right. }  \notag \\
&&-\frac{1}{2}\sqrt{\left.
\begin{array}{c}
\left( 2p-(\text{$c_{3}$}+1)(2p-1)\cos ^{2}(r)\right)  \\
\left( (\text{$c_{3}$}+1)(2p-1)\cos ^{2}(r)-2p+2\right)
\end{array}%
\right. }  \notag \\
&&
\end{eqnarray}%
where the super-scripts AD, Dep and BF correspond to amplitude damping,
depolarizing and bit flip channels respectively. The results are consistent
with Refs. [46, 47] and can be easily verified from the expressions
(equations 22-24) by setting $r=\mu =0$ and $\mu =0$ respectively.

\section{Discussions}

Analytical expressions for the concurrence are calculated for $X$-type
initial state in non-inertial frames influenced by amplitude damping,
depolarizing and bit flip channels. In figures 1 and 2, the concurrence is
plotted as a function of memory parameter $\mu $ for $p=0.3$ and $p=0.7$\
respectively, for amplitude damping, depolarizing and bit flip channels. The
first panel (column-wise) corresponds to Bell states, whereas the second and
third panels correspond to Werner and general initial states, respectively.
It is seen that for Bell basis states, entanglement sudden death can be
avoided in case of amplitude damping and depolarizing channels in the
presence of correlated noise. However, it is possible to fully avoid ESD for
all the channels under consideration for $\mu >0.75$ (which can be seen from
the figure). On the other hand, ESD can also be avoided for Werner and
general initial states as well in case of amplitude damping channel at
higher degree of correlations. As the value of acceleration $r$ increases$,$
the entanglement degradation is enhanced which is more prominent for lower
range of memory parameter $\mu $. It is also seen that bit-flip noise
heavily influences the entanglement of the system as compared to the
amplitude damping and depolarizing channels for lower level of decoherence.
Whereas at higher level of decoherence, damping effect of amplitude damping
channel becomes more prominent (see figure 2). Furthermore, it can be seen
that initial state plays an important role in the system-environment
dynamics of entanglement in non-inertial frames.

In figure 3, the concurrence is plotted as a function of decoherence
parameter $p$\ for $\mu =0.5$\ for amplitude damping, depolarizing and bit
flip channels. It can be seen that the concurrence is heavily damped by
different environments. This effect is much prominent for amplitude damping
and bit flip channels. It is seen that maximum entanglement degradation
occurs at $p=0.5$ in case of bit flip channel and entanglement rebound
process take place for $p>0.5$. Furthermore, no ESD behaviour is seen for
depolarizing channel in the presence of correlated noise. The degree of
entanglement degradation enhances as one shifts from Bell type initial state
to the case of general initial state. However, it saturates for infinite
acceleration limit $(r=\pi /4)$.

In figures 4 and 5, the concurrence is plotted as a function of Rob's
acceleration $r$\ and decoherence parameter $p$ with $\mu =0.3$ and $0.7$
for amplitude damping, depolarizing and bit flip channels. The upper panel
(row-wise) corresponds to Bell states, whereas the middle and lower panels
correspond to Werner and general initial states respectively. From figure 5,
it can be seen that ESD can be fully avoided for all the three types of
initial states under depolarizing and bit flip noises at 50\% quantum
correlations. On the other hand, ESD behaviour is seen only in case of
amplitude damping channel for $p>0.75$. Therefore, different environments
affect the entanglement of the system differently. In figure 6, the
concurrence is plotted as a function of memory parameter $\mu $\ and
decoherence parameter $p$ with $r=\pi /4$ for amplitude damping,
depolarizing and bit flip channels. It is shown that maximum ESD occurs in
case of amplitude damping channel. However, ESD can be avoided for
depolarizing channel for $\mu >0$ even for maximum value of decoherence i.e.
$p=1.$

\section{Conclusions}

Environmental effects on the entanglement dynamics of Dirac fields in
non-inertial frames is investigated by considering $X$-type initial state
shared between the two partners. It is assumed that the Rob is in
accelerated frame moving with uniform acceleration whereas Alice is the
stationary observer. The concurrence is used to investigate the decoherence
and correlated noise effects on the entanglement of the system. Different
initial states are considered such as Bell basis, Werner type and general
initial states. It is seen that in case of Bell basis states, entanglement
sudden death can be avoided for amplitude damping and depolarizing channels
in the presence of correlated noise. Whereas, for Werner like and general
initial states, the entanglement sudden death occurs more rapidly as the
value of decoherence parameter $p$ and Rob's acceleration $r$ increase.
Therefore, the initial state plays an important role in the
system-environment dynamics of entanglement in non-inertial frames. It is
shown that bit-flip channel and amplitude damping channels heavily influence
the entanglement of the system as compared to the depolarizing channel. It
is possible to avoid ESD for all the channels under consideration for $\mu
>0.75$ irrespective of the type of initial state considered. The effect of
environment is much stronger than that of Rob's acceleration $r$.
Furthermore, no ESD occurs for depolarizing channel for any value of $p$ and
$r$, in the presence of correlated noise. In conclusion, correlated noise
compensates the loss of entanglement caused by the Unruh effect.

{\huge Figures captions}\newline
\textbf{Figure 1}. (Color online). The concurrence is plotted as a function
of memory parameter $\mu $\ for $p=0.3$\ for amplitude damping, depolarizing
and bit flip channels, where the abbreviations AD, Dep and BF correspond to
amplitude damping, depolarizing and bit flip channels respectively.\newline
\textbf{Figure 2}. (Color online). The concurrence is plotted as a function
of memory parameter $\mu $\ for $p=0.7$\ for amplitude damping, depolarizing
and bit flip channels.\newline
\textbf{Figure 3}. (Color online). The concurrence is plotted as a function
of decoherence parameter $p$\ for $\mu =0.5$\ for amplitude damping,
depolarizing and bit flip channels.\newline
\textbf{Figure 4}. (Color online). The concurrence is plotted as a function
of Rob's acceleration, $r$\ and decoherence parameter, $p$ with $\mu =0.3$
for amplitude damping, depolarizing and bit flip channels.\newline
\textbf{Figure 5}. (Color online). The concurrence is plotted as a function
of Rob's acceleration, $r$\ and decoherence parameter, $p$ with $\mu =0.7$
for amplitude damping, depolarizing and bit flip channels.\newline
\textbf{Figure 6}. (Color online). The concurrence is plotted as a function
of memory parameter, $\mu $\ and decoherence parameter, $p$ with $r=\pi /4$
for amplitude damping, depolarizing and bit flip channels.\newline
{\Huge Table Caption}\newline
\textbf{Table 1}. Single qubit Kraus operators for amplitude damping,
depolarizing and bit flip channels where $p$ represents the decoherence
parameter.\newline
\begin{figure}[tbp]
\begin{center}
\vspace{-2cm} \includegraphics[scale=0.8]{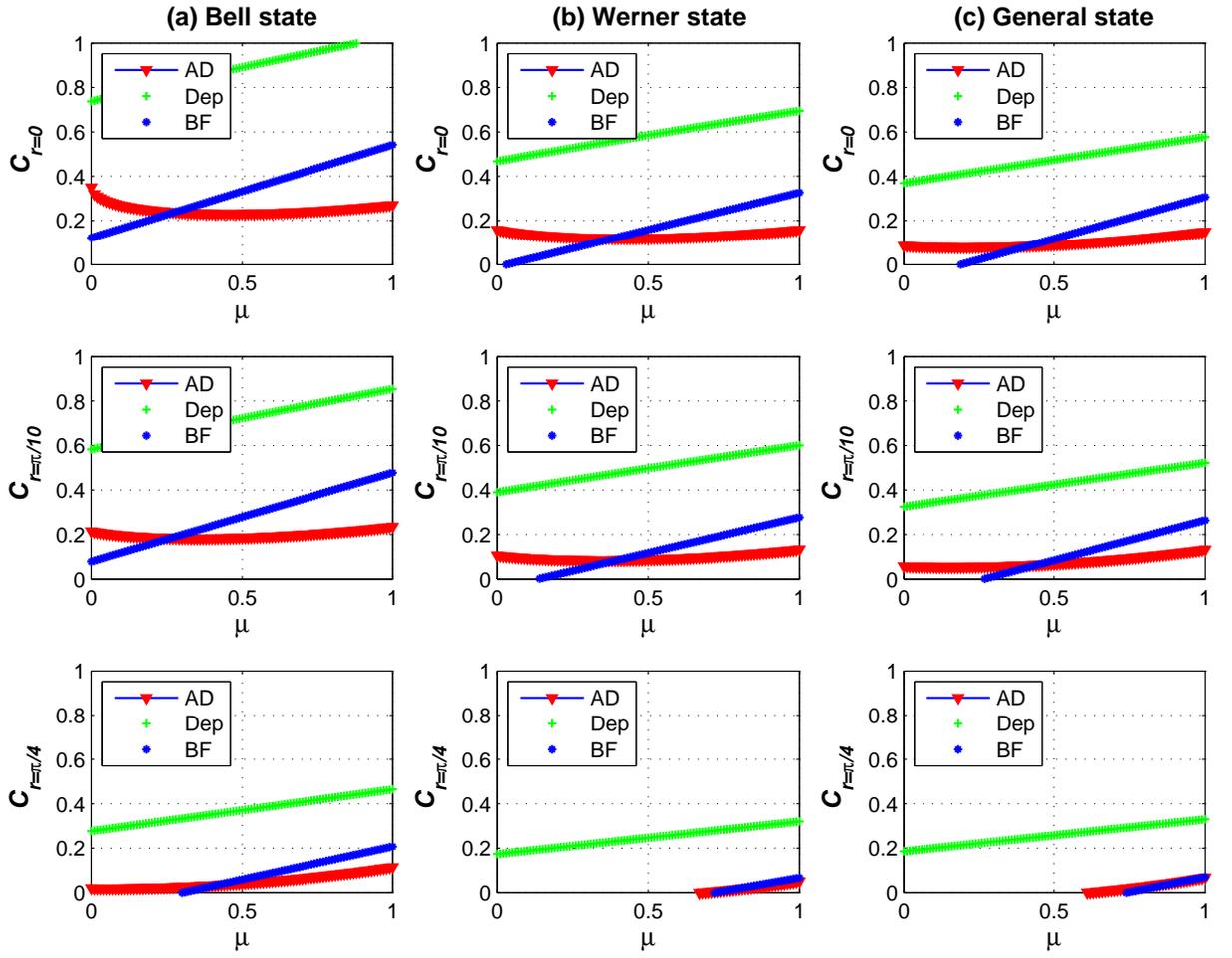} \\[0pt]
\end{center}
\caption{(Color online). The concurrence is plotted as a function of memory
parameter $\protect\mu $\ for $p=0.3$\ for amplitude damping, depolarizing
and bit flip channels.}
\end{figure}

\begin{figure}[tbp]
\begin{center}
\vspace{-2cm} \includegraphics[scale=0.8]{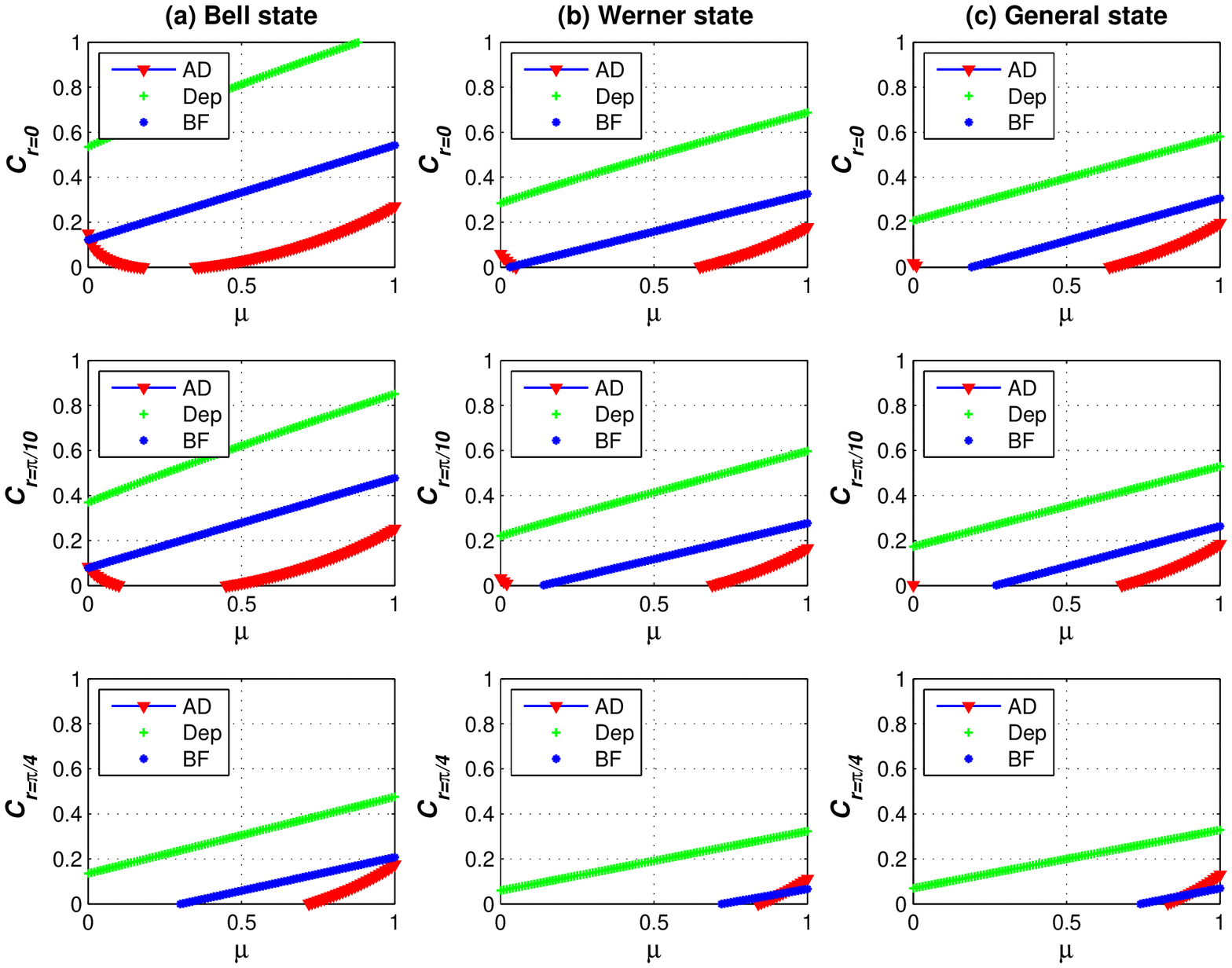} \\[0pt]
\end{center}
\caption{(Color online). The concurrence is plotted as a function of memory
parameter $\protect\mu $\ for $p=0.7$\ for amplitude damping, depolarizing
and bit flip channels.}
\end{figure}

\begin{figure}[tbp]
\begin{center}
\vspace{-2cm} \includegraphics[scale=0.8]{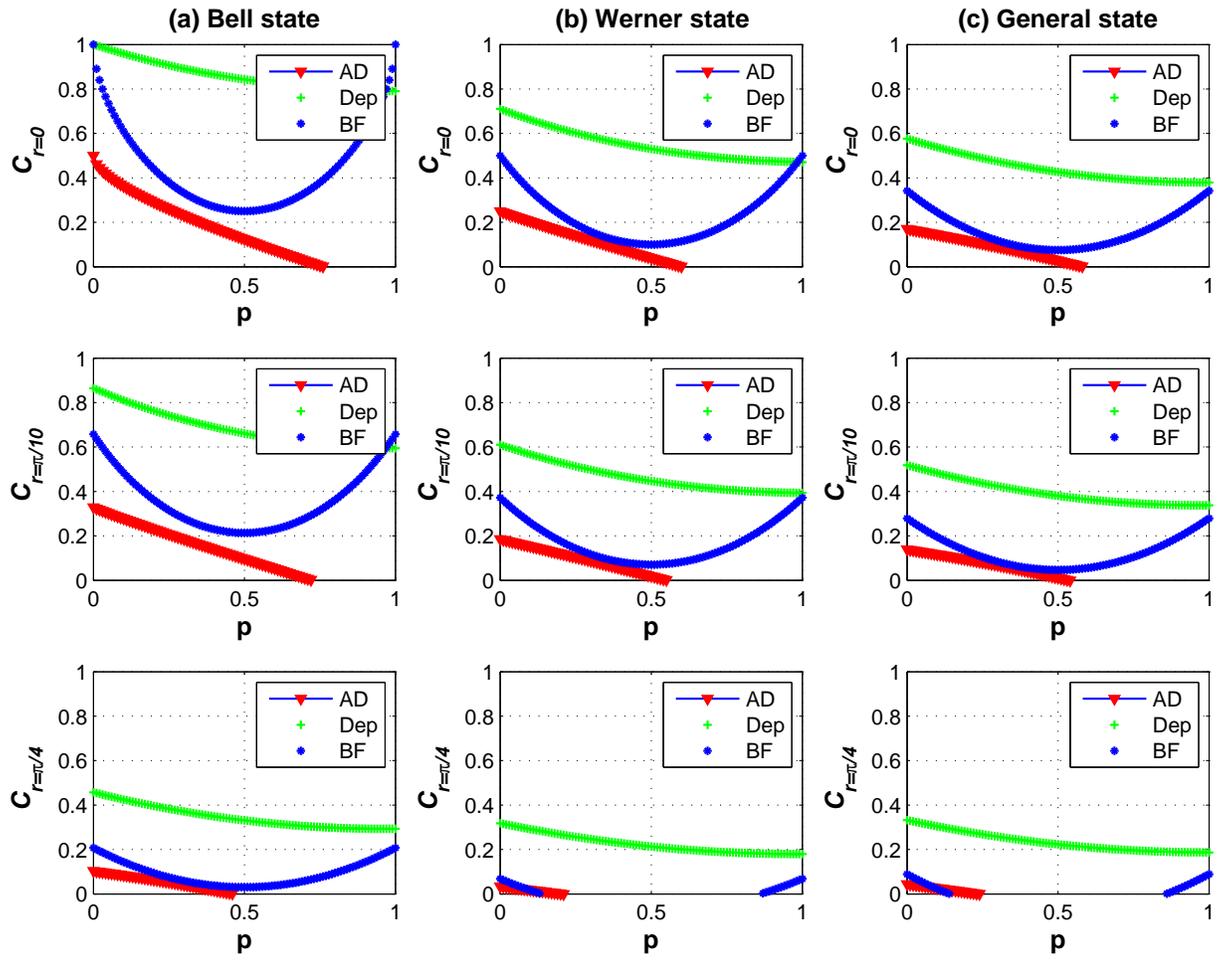} \\[0pt]
\end{center}
\caption{(Color online). The concurrence is plotted as a function of
decoherence parameter $p$\ for $\protect\mu =0.5$\ for amplitude damping,
depolarizing and bit flip channels.}
\end{figure}

\begin{figure}[tbp]
\begin{center}
\vspace{-2cm} \includegraphics[scale=0.8]{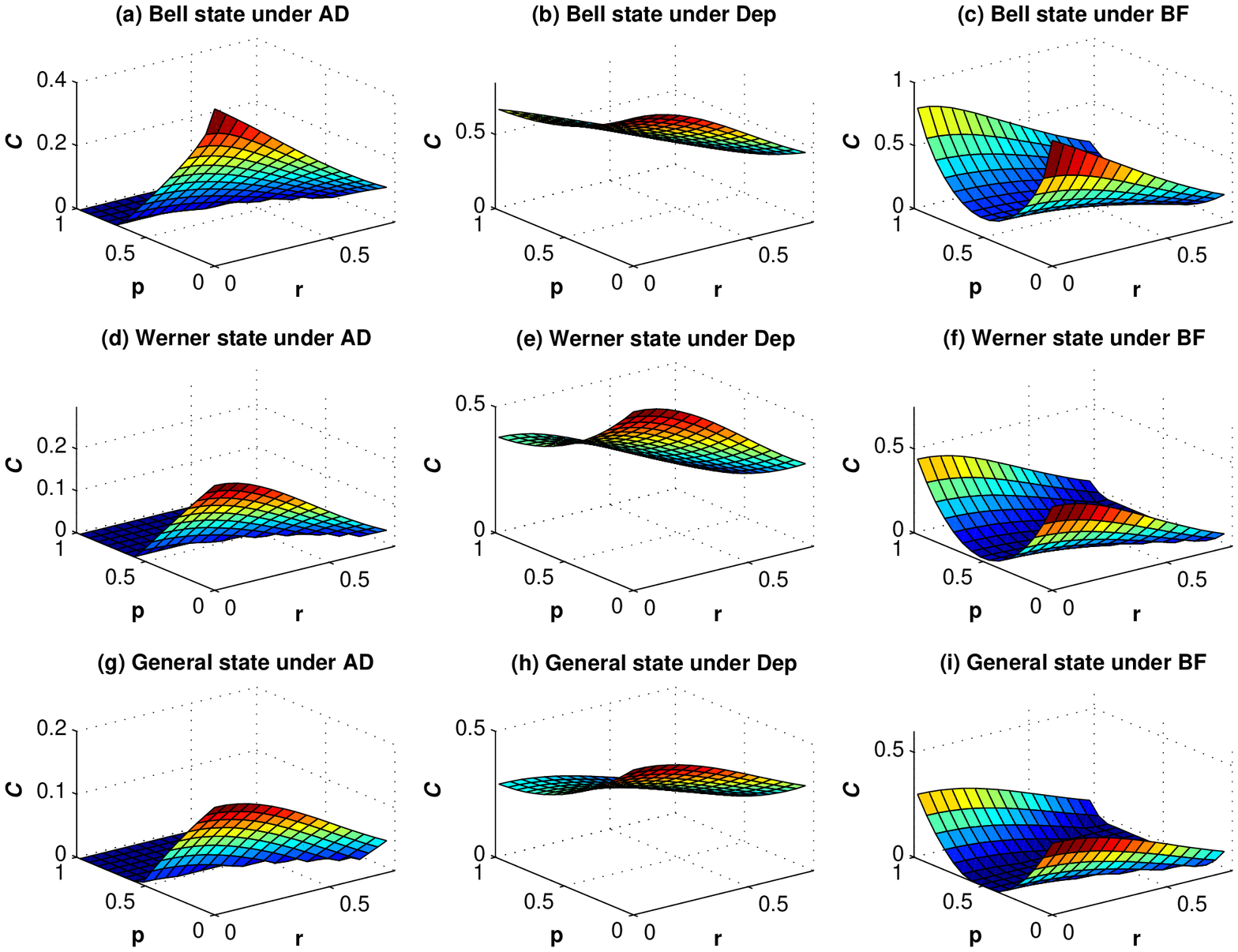} \\[0pt]
\end{center}
\caption{(Color online). The concurrence is plotted as a function of Rob's
acceleration, $r$\ and decoherence parameter, $p$ with $\protect\mu =0.3$
for amplitude damping, depolarizing and bit flip channels.}
\end{figure}

\begin{figure}[tbp]
\begin{center}
\vspace{-2cm} \includegraphics[scale=0.8]{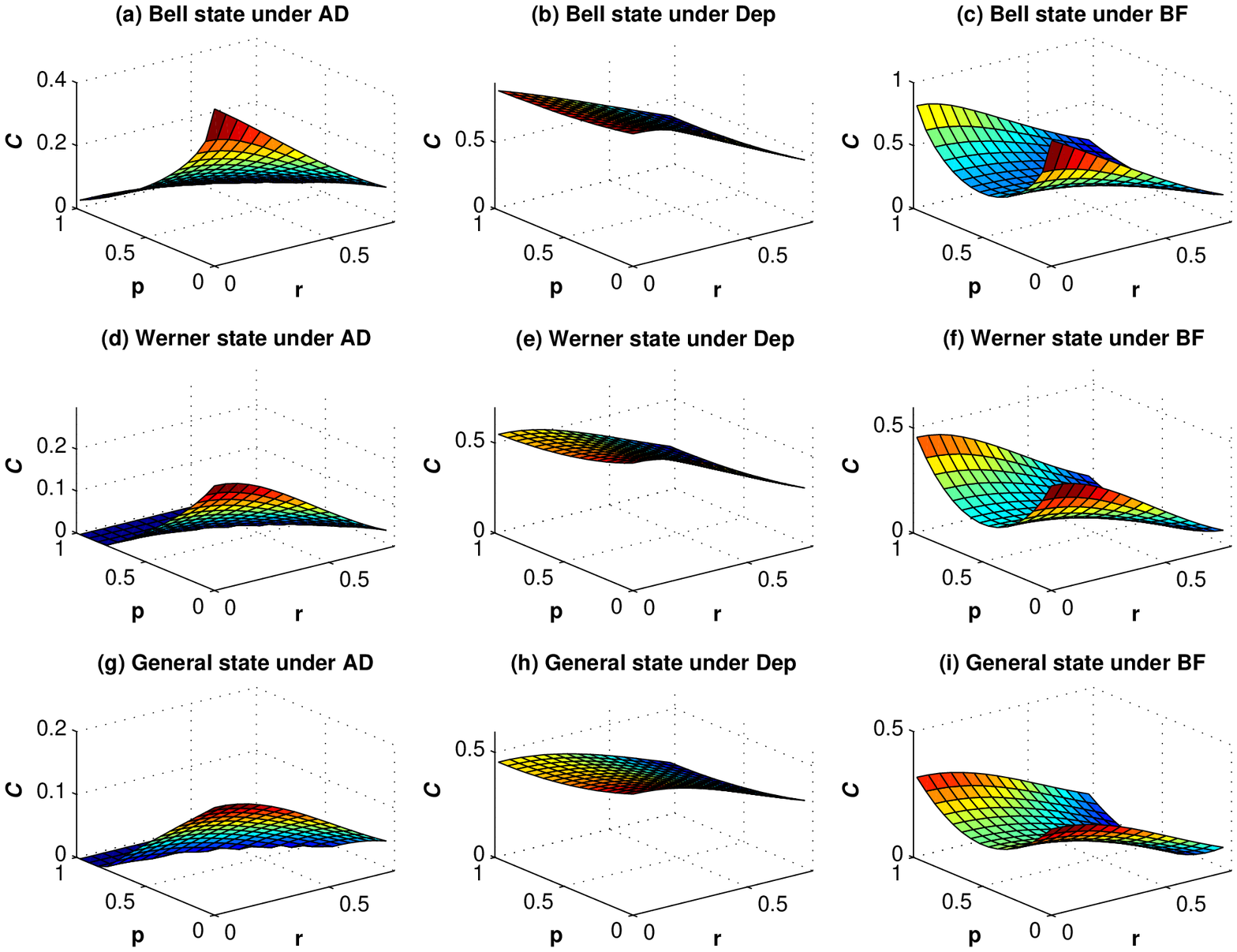} \\[0pt]
\end{center}
\caption{(Color online). The concurrence is plotted as a function of Rob's
acceleration, $r$\ and decoherence parameter, $p$ with $\protect\mu =0.7$
for amplitude damping, depolarizing and bit flip channels.}
\end{figure}

\begin{figure}[tbp]
\begin{center}
\vspace{-2cm} \includegraphics[scale=0.8]{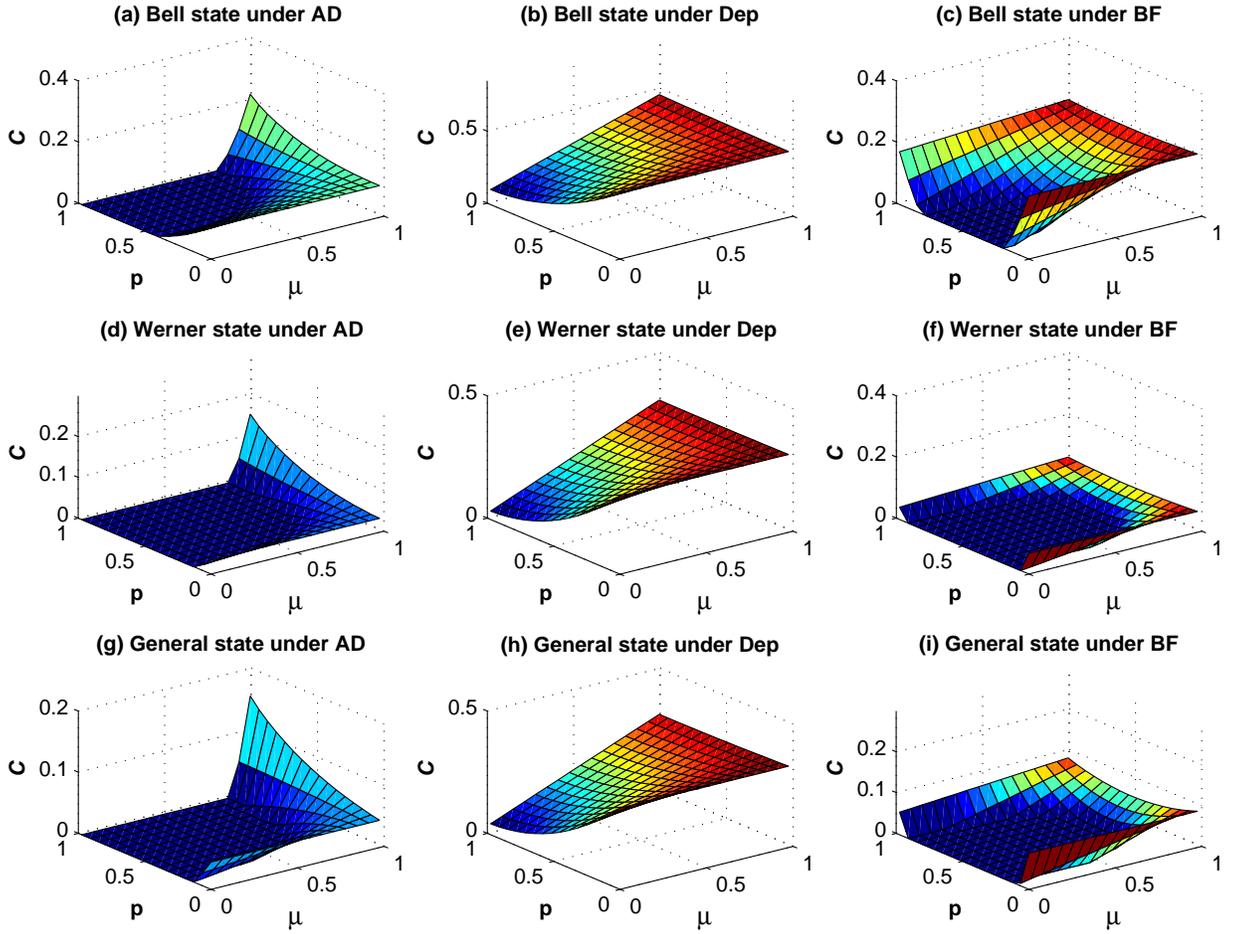} \\[0pt]
\end{center}
\caption{(Color online). The concurrence is plotted as a function of memory
parameter, $\protect\mu $\ and decoherence parameter, $p$ with $r=\protect%
\pi /4$ for amplitude damping, depolarizing and bit flip channels.}
\end{figure}
\begin{table}[tbh]
\caption{Single qubit Kraus operators for amplitude damping, depolarizing
and bit flip channels where $p$ represents the decoherence parameter.}$%
\begin{tabular}{|l|l|}
\hline
&  \\
$\text{Amplitude damping channel}$ & $A_{0}=\left[
\begin{array}{cc}
1 & 0 \\
0 & \sqrt{1-p}%
\end{array}%
\right] ,$ $A_{1}=\left[
\begin{array}{cc}
0 & \sqrt{p} \\
0 & 0%
\end{array}%
\right] $ \\
&  \\ \hline
&  \\
Depolarizing$\text{ channel}$ & $%
\begin{tabular}{l}
$A_{0}=\sqrt{1-\frac{3p}{4}I},\quad A_{1}=\sqrt{\frac{p}{4}}\sigma _{x}$ \\
$A_{2}=\sqrt{\frac{p}{4}}\sigma _{y},\quad \quad $\ $\ A_{3}=\sqrt{\frac{p}{4%
}}\sigma _{z}$%
\end{tabular}%
$ \\
&  \\ \hline
&  \\
Bit flip$\text{ channel}$ & $A_{0}=\sqrt{1-p}I,A_{1}=\sqrt{p}\sigma _{x}$ \\
&  \\ \hline
\end{tabular}%
$%
\label{di-fit}
\end{table}

\end{document}